\documentclass[12pt]{article}
\usepackage{amssymb,amsmath,epsfig,makecell,listings, amsthm, mathtools, paralist, enumitem}
\usepackage{setspace}
\usepackage{cite}
\usepackage{url}

\usepackage[font=small, labelfont=bf]{caption}
\usepackage[font=footnotesize, labelfont=bf]{subcaption}
\usepackage{capt-of}

\DeclareCaptionFont{singlespacing}{\setstretch{1}}
\usepackage{color}

\theoremstyle{plain}

\theoremstyle{definition}

\newcommand{\N}{\ensuremath{\mathbb{N}}}

\title{The Topological ``Shape'' of Brexit}
\author{Bernadette J. Stolz\footnote{Mathematical Institute, University of Oxford}, Heather A. Harrington\footnote{Mathematical Institute, University of Oxford}, and Mason A. Porter\footnote{Department of Mathematics, UCLA; Mathematical Institute, University of Oxford; and CABDyN Complexity Centre, University of Oxford}}
\date{\today}

\begin{document}

\maketitle

\section*{Abstract}

Persistent homology is a method from computational algebraic topology that can be used to study the ``shape'' of data. We illustrate two filtrations --- the weight rank clique filtration and the Vietoris--Rips (VR) filtration --- that are commonly used in persistent homology, and we apply these filtrations to a pair of data sets that are both related to the 2016 European Union ``Brexit'' referendum in the United Kingdom. These examples consider a topical situation and give useful illustrations of the strengths and weaknesses of these methods.


 \section{Introduction}
  
\emph{Persistent homology} (PH)~\cite{Edelsbrunner2008,Edelsbrunner2010,Ghrist2008,Carlsson2009,Ghrist2014,otter2015} is a method from computational algebraic topology that can be used to study the ``shape'' of data. The studied data typically consists of networks or high-dimensional point clouds. PH is concerned with topological invariants, such as connectedness and holes in high-dimensional objects, over a range of different scales. Topological features that persist over many scales are often considered to be significant for data shape, and features that persist only over a small range of scales are construed as noise. PH has been applied to study an increasingly diverse set of phenomena in a large variety of subjects. A small set of the myriad examples include granular materials~\cite{kramar2013}, protein binding sites~\cite{Kovacev2015}, brain-artery trees~\cite{Bendich2014}, and functional brain networks \cite{Curto2016,Giusti2016,stolz2016}. 

In the present article, we illustrate two filtrations that are commonly used in PH: the weight rank clique filtration (WRCF)~\cite{Petri2013} and the Vietoris--Rips (VR) filtration~\cite{Ghrist2008}. We apply these filtrations to a pair of data sets that are both related to the recent European Union (EU) referendum in the United Kingdom (UK) \cite{brexit-wikipedia}. The first data set consists of a weighted network based on the year that each European country joined the EU and the second data set consists of two point clouds that were constructed from voting results of the UK referendum. We give a detailed description of both data sets in Subsections~\ref{sec: Network} and~\ref{sec: Point cloud} respectively.

    
\section{The EU Network} 
  
\subsection{Constructing a Network}\label{sec: Network}
    
\begin{figure}[h!]
\centering\includegraphics[width=\textwidth]{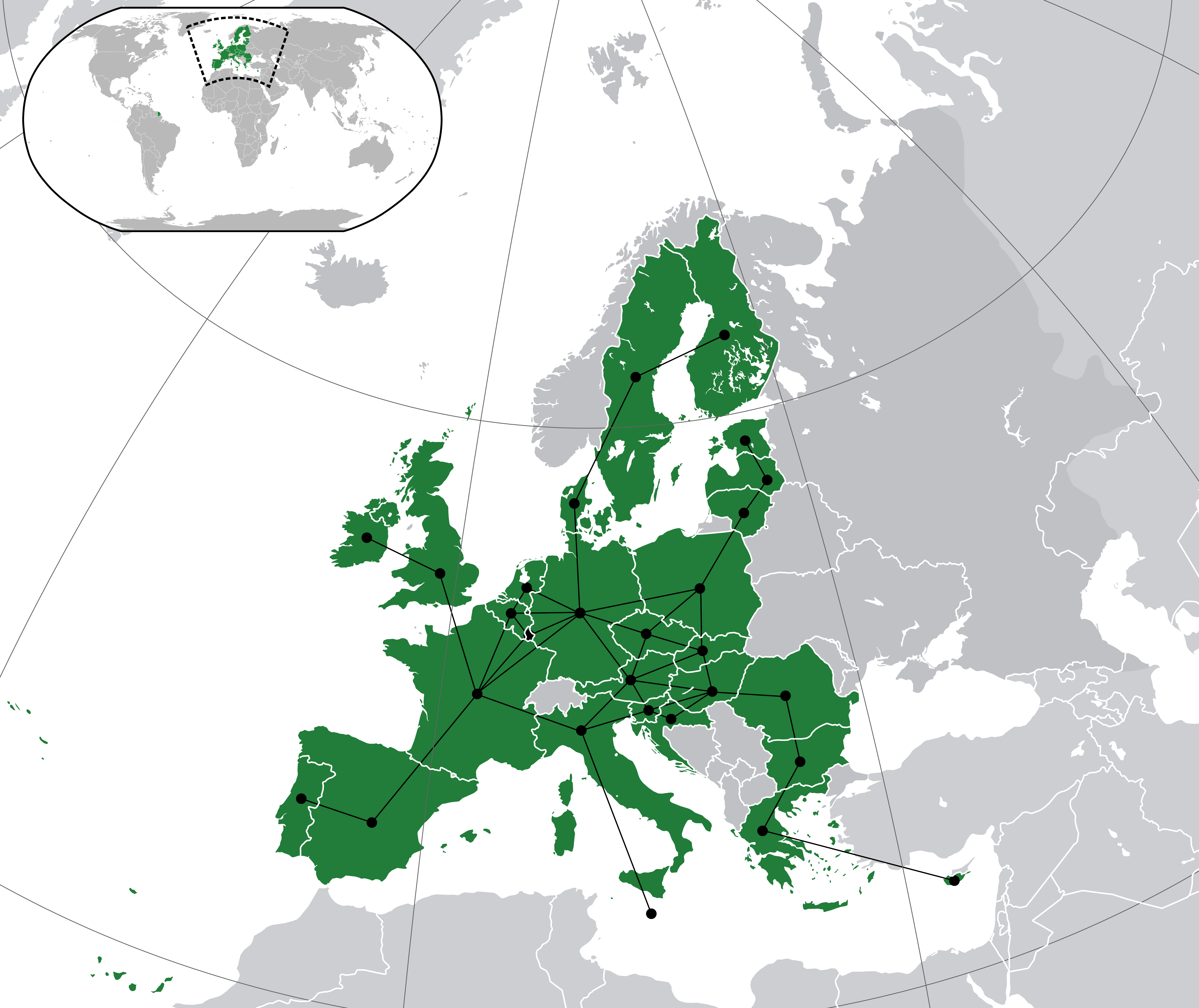}
\caption{Network of EU countries. We connect two countries by an edge if they are considered neighbors via a border (either in Europe or abroad), a bridge, or a tunnel. We define the edge weight between two countries to be the later of the two years in which they joined the EU. [We obtained the employed map from Wikimedia \cite{eu-wiki}.]
}\label{fig:objects}
\end{figure}

We define a network of EU countries based on the geographical location of the countries and the year in which they joined the European Union. Each current EU country (as of June 2016) is a node in the network. We connect two countries by an edge if they are neighbors via a common border (either in Europe or via overseas territories), a bridge (Denmark--Sweden), or a tunnel (UK--France). We also connect Malta to Italy and connect Cyprus to Greece because of their geographical proximity. We show the resulting network in Fig.~\ref{fig:objects}. We define the edge weight $w_{i,j}$ between countries $i$ and $j$ to be the later of the two years in which those two countries joined the EU. That is, 
\begin{equation*}
	w_{i,j} = \max{\{\text{year in which country } i \text{ joined the EU},\text{year in which country } j \text{ joined the EU}\}}\,.
\end{equation*}

 \subsection{Weight Rank Clique Filtration}
  
We use a WRCF~\cite{Lee2012,Petri2013} on the above weighted network to obtain a sequence of embedded graphs, on which we examine PH. One constructs a WRCF as follows:
\begin{enumerate}
\item Define filtration step $0$ as the set of all nodes.
\item Rank all edge weights $\{ \nu_{1},\dots, \nu_{\text{end}} \}$, with $\nu_{1} = \max_{1 \leq i,j  \leq 28}{w_{i,j}}$ and $\nu_{\text{end}} = \min_{1 \leq i,j  \leq 28}{w_{i,j}}$. 
\item In filtration step $t$, threshold the graph at weight $\nu_t$  
to create a binary (i.e., unweighted) graph. After thresholding, edges with weight at least $\nu_t$ are present, and edges with smaller weights are absent.
\item Find all maximal $c$-cliques for $c \in \N$, and define them to be $c$-simplices.
\end{enumerate}

\begin{figure}[h!]
\centering
\centering\includegraphics[width=\textwidth]{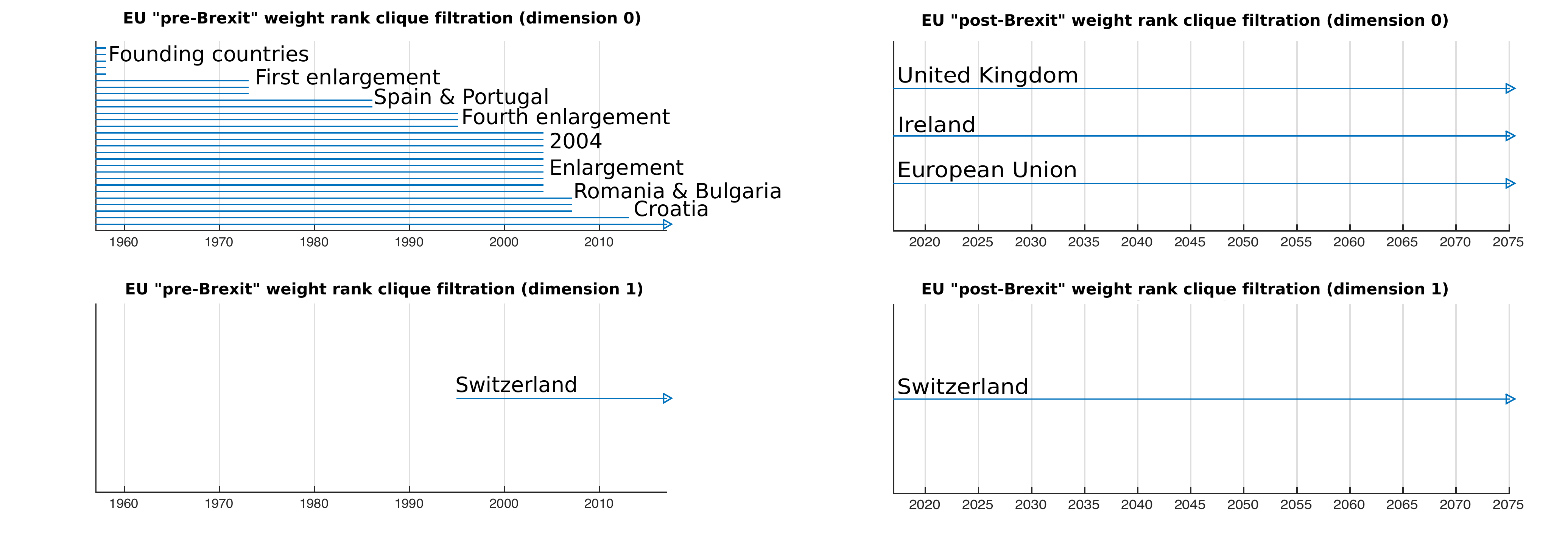}
\caption{Barcodes for dimensions $0$ (connected components) and $1$ (loops) obtained from a weight rank clique filtration performed on current and hypothetical future EU networks. See Table \ref{tab:EUEnlargements} for a listing of EU enlargements.
}\label{fig:WRCFBarcodes}

\captionof{table}{EU enlargements listed in the barcodes.}
\begin{tabular}{ | l | p{11cm} | }
  \hline
 Founding countries & Belgium, Germany, France, Italy, Luxembourg, and The Netherlands \\
  \hline
First enlargement & Denmark, Ireland, and the United Kingdom \\
 \hline
  Fourth enlargement  & Austria, Finland, and Sweden\\
   \hline
   2004 enlargement &  Czech Republic, Estonia, Cyprus, Latvia, Lithuania, Hungary, Malta, Poland, Slovakia, and Slovenia\\
    \hline
\end{tabular}
\label{tab:EUEnlargements}
\end{figure}

We perform the WRCF on the EU network before the EU referendum and on a network based on the future EU after the referendum results are implemented. In this future EU, we have removed the two edges between the UK and the other EU countries as if they never existed in the first place, but we keep all nodes as before. We run the filtration from years 1953--2016 for the current EU network and 2016 onwards for the hypothetical future EU network. To visualize the results of the PH computations, we show barcode diagrams \cite{Ghrist2008} in Fig.~\ref{fig:WRCFBarcodes}.

In dimension $0$, PH detects the connected components of the network in every filtration step. In our case, every bar in the 0-dimensional barcode represents one EU country. The countries merge as the EU grows over time until there is one large EU component in 2013. Because of how we construct the edge weights in the network, we note that although Greece joined the EU in 1981, it only joins the large EU component in 2007 when its neighbor Bulgaria joined the EU. Similarly, in the barcode after the 2016 EU referendum, both the UK and Ireland are in separate components because the latter (in our hypothetical ``post-Brexit" world) has lost its connection to the main EU component via the UK even though it is still a member of the EU.

In dimension $1$, PH detects loops in the network. In the case of the WRCF these loops consist of at least four edges. In other words, edges between three pairwise neighboring countries are not registered as loops, but the loop created by edges between Switzerland's neighboring countries of leads to a bar in the $1$-dimensional barcode starting in 1995 when Austria joins the EU. We remark that this is also the only occurrence of a loop in this network.


\section{The Voting Point Cloud}
  
\subsection{Data Acquisition}\label{sec: Point cloud}
  
\begin{figure}[h!]
  \centering\includegraphics[width=.55\textwidth]{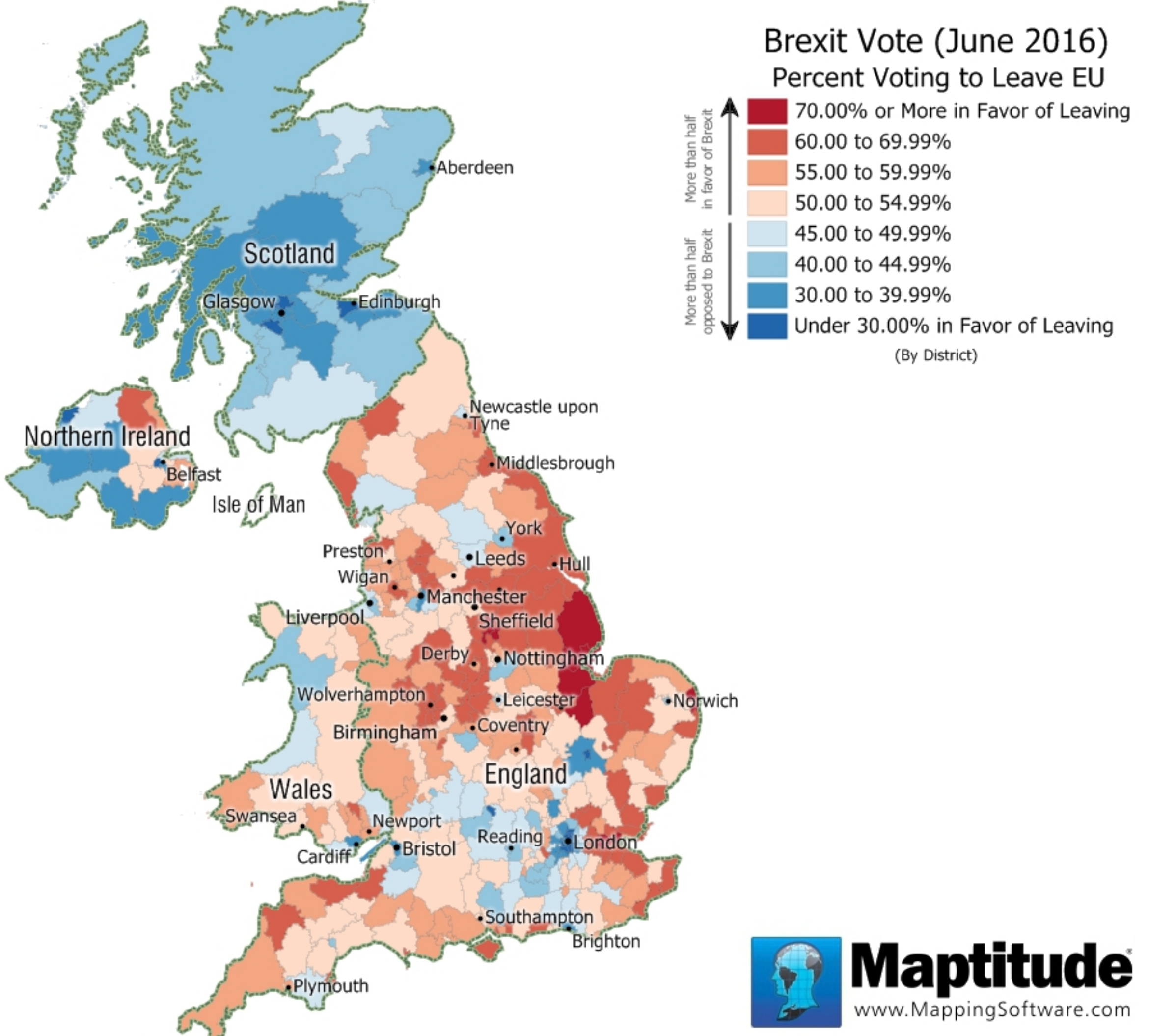}
\centering\includegraphics[width=.4\textwidth]{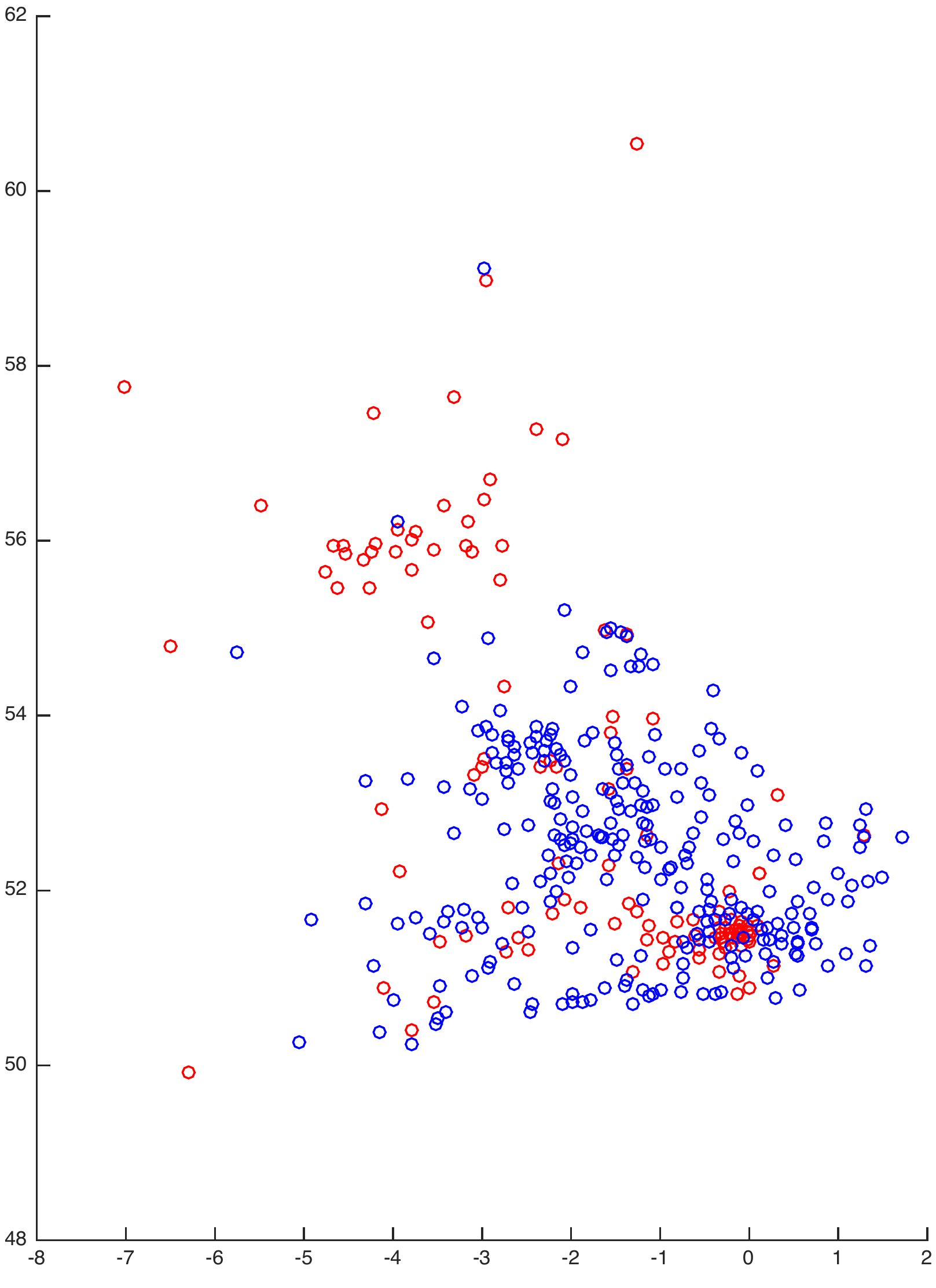}
\caption{Point clouds based on the 2016 EU referendum voting result in the UK. We show the coordinates of ``leave'' districts in blue and the coordinates of ``remain'' districts in red.
}\label{fig:objects2}
\end{figure}

We obtained UK voting data via the website ``Number Cruncher Politics'' \cite{ncp}. We extracted coordinates for one city per voting district by combining python scripts and the Google Maps API. We thereby construct two point clouds: one, which we call the \emph{leave point cloud}, contains the coordinates of cities in voting districts that voted to leave the EU; the other, which we call the \emph{remain point cloud}, contains the coordinates of cities in voting districts that voted to stay in the EU. We show both point clouds in Fig.~\ref{fig:objects2}.

 \subsection{Vietoris--Rips Filtration}

\begin{figure}[h!]
\centering\includegraphics[width=.45\textwidth]{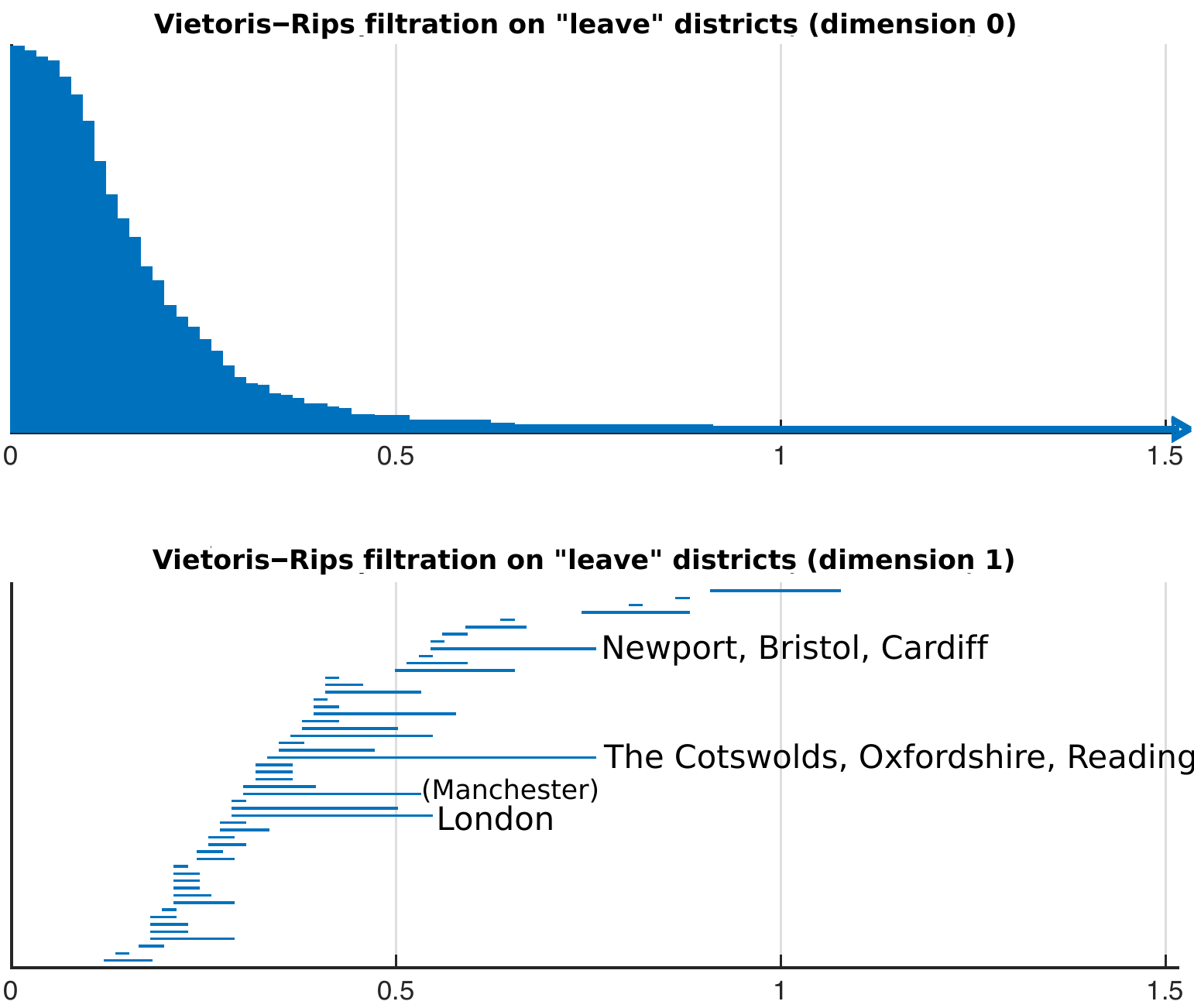}
\centering\includegraphics[width=.45\textwidth]{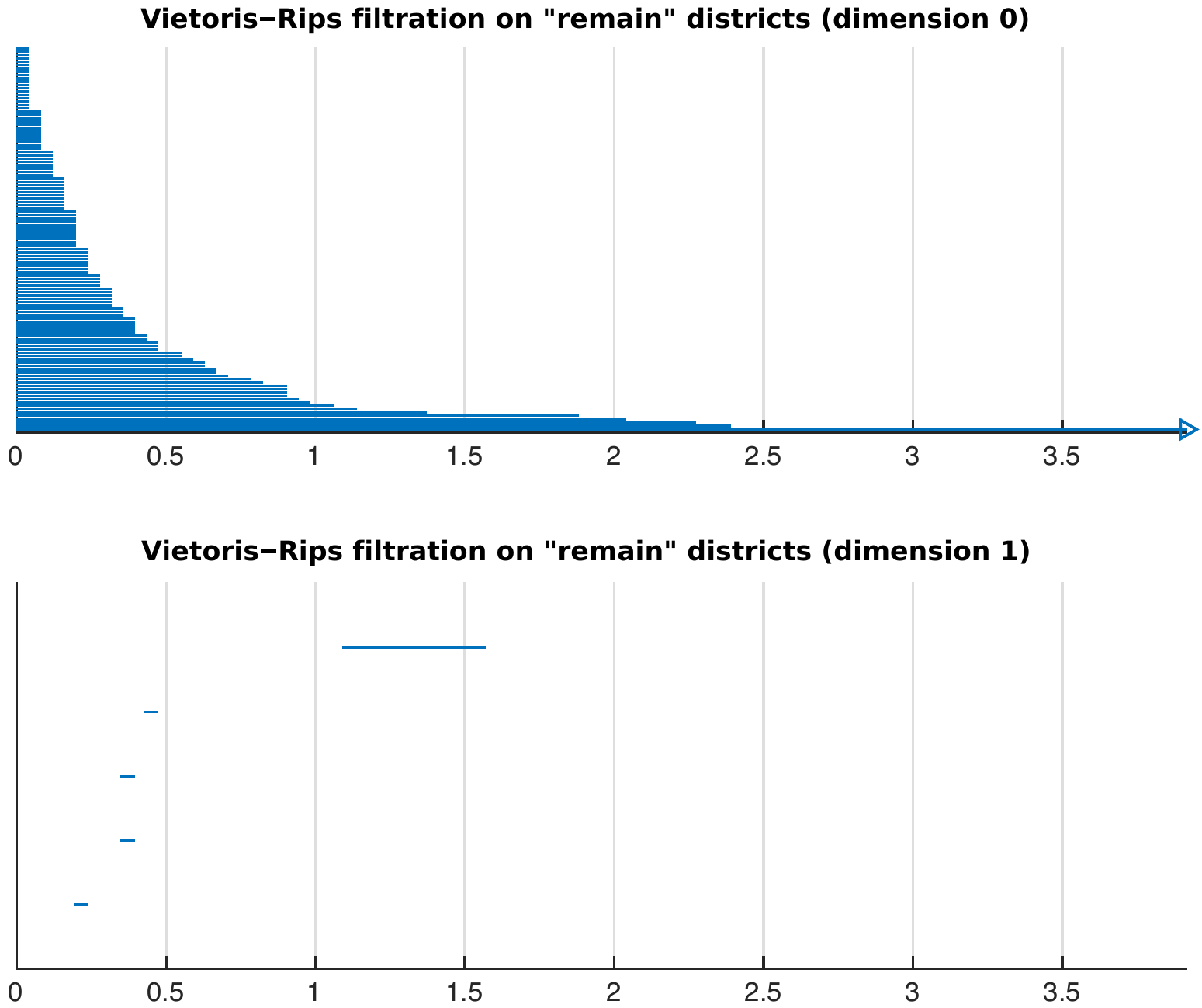} \\
\centering\includegraphics[width=.45\textwidth]{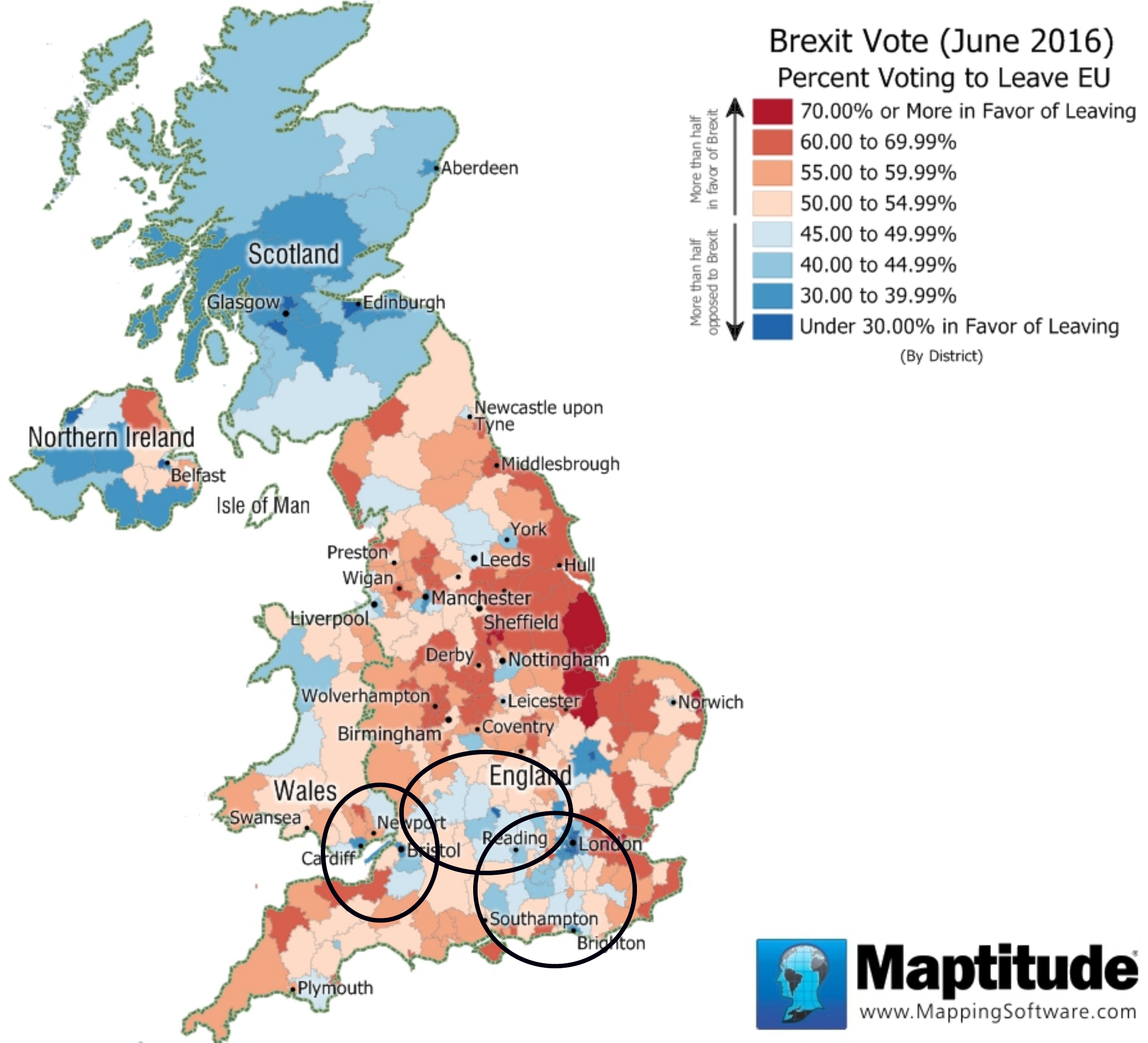}
\centering\includegraphics[width=.45\textwidth]{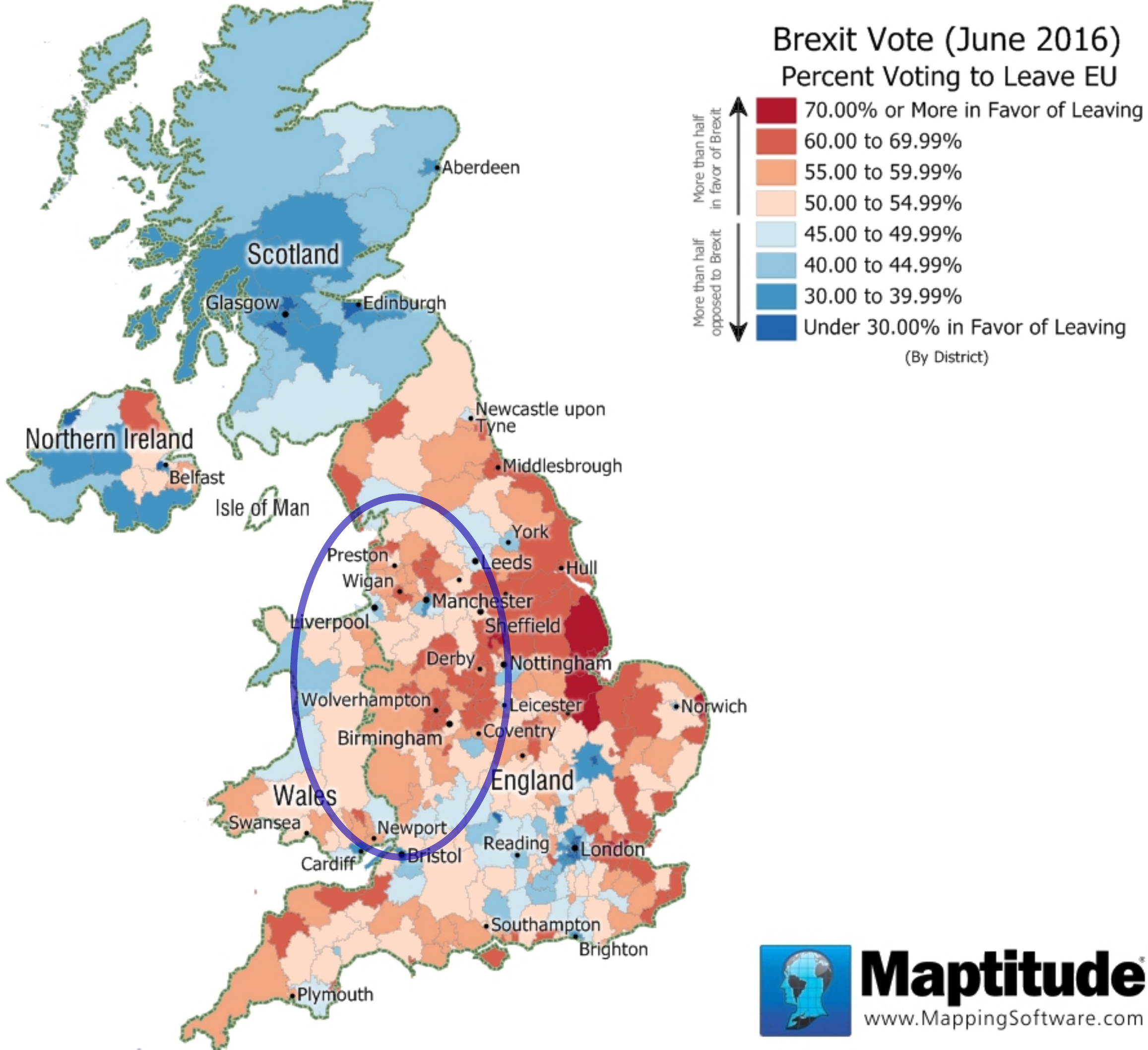}
\caption{(Top two rows) Barcodes for dimensions $0$ and $1$ from a Vietoris--Rips filtration on (left) the leave point cloud and (right) the remain point cloud. (Bottom row) UK referendum voting map including the loops in dimension $1$ of the Vietoris--Rips filtration.
}\label{fig:objects3}
\end{figure}

We apply a VR filtration~\cite{Ghrist2008} to our point-cloud data. One constructs a VR filtration as follows:

 \begin{enumerate}
\item Choose a sequence $\epsilon$ of increasing distances: $\epsilon = \{ \epsilon_1, \dots, \epsilon_n \}$.
\item In the $i$th filtration step, define $k$-simplices using unordered $(k+1)$-tuples whose pairwise distance is at most $\epsilon_i$.
\end{enumerate}
We show the resulting barcodes in Fig.~\ref{fig:objects3}.

Our calculations reveal multiple loops in dimension $1$ of the leave district barcodes that are more persistent than other features of the barcode. In particular, London and the region around Oxfordshire, Reading, and The Cotswolds stand out as two holes in the leave district point cloud, as they voted in the opposite way as their surrounding voting districts. However, we also note that the VR filtration detects a persistent hole around Manchester, where there ought to be a non-persistent hole. This is due to the fact that the points representing the surrounding voting districts are located far away from Manchester, and this leads to an artificially large hole. In the remain point cloud, we find a persistent hole around Birmingham in dimension $1$.

  
\section{Conclusions}

Our two examples illustrate the strengths and limitations of the applied PH methods. For the weighted rank clique filtration on the country network, we find a single large (and relevant) network loop around Switzerland in dimension $1$. The WRCF in dimension $0$ gives a good idea of the years when countries join the EU, but the results also depend heavily on the neighbors in the network, which leads to the method suggesting erroneously that Greece joined the EU later than it actually did and Ireland being illustrated as a separate component in the EU network after the UK leaves the EU.

The Vietoris--Rips filtration points towards large groups of regions that have voted differently than their surrounding regions, but we also find that the holes in dimension $1$ depend on the choice of points to represent the voting districts.

The two methods that we have illustrated are of course not the only topological methods that one could apply to the data set. For example, one could build a single filtration for the EU network over all years that allows edges to disappear from one filtration step to the next by using zigzag persistent homology~\cite{Carlsson2010, Tausz2011}, or one could focus on counting the number of persistent loops in the point-cloud data and compare this to point clouds of other UK votes to see whether there are recurring vote shapes (which could represent recurring voting patterns).

  
\section{Acknowledgements}

We thank Matt Singh from Number Cruncher Politics for providing us with the referendum voting data. We also thank Joshua Bull for his help with Python webcrawling and Jared Tanner for helpful discussions. BJS also gratefully acknowledges the EPSRC and MRC (grant number EP/G037280/1), and she thanks F. Hoffmann--La Roche AG for funding her doctoral studies. HAH acknowledges funding from EPSRC Fellowship EP/K041096/1.


\end{document}